\documentclass[a4paper,12pt]{article}

\usepackage[top=3cm,left=3cm,bottom=3cm,right=3cm]{geometry}
\usepackage{amsmath}
\usepackage{graphicx}
\usepackage[english,frenchb]{babel}
\usepackage{times}
\usepackage[sort&compress]{natbib}
\usepackage[usenames,dvipsnames]{color}
\usepackage{amssymb}
\usepackage{epstopdf}
\usepackage{enumerate}
\usepackage{wrapfig}
\usepackage{multicol}
\usepackage{eurosym}
\usepackage{pdfpages}
\usepackage{tabularx}
\usepackage{booktabs}
\usepackage{colortbl}
\usepackage{makeidx}
\usepackage{caption}
\usepackage[utf8]{inputenc}

\usepackage[T1]{fontenc}
\usepackage{setspace}
\usepackage{fancyhdr}

\usepackage{titlesec}

\titleformat{\paragraph}
{\normalfont\normalsize\bfseries}{\theparagraph}{1em}{}
\titlespacing*{\paragraph}
{0pt}{3.25ex plus 1ex minus .2ex}{1.5ex plus .2ex}

\usepackage[bookmarks, colorlinks, breaklinks, pdftitle={David Chavalarias},pdfauthor={David Chavalarias}]{hyperref}  
\hypersetup{linkcolor= MidnightBlue,citecolor= MidnightBlue,filecolor=black,urlcolor= MidnightBlue}

\newcommand{\tmb}{\textcolor{MidnightBlue}}

\makeatletter
\renewcommand{\section}{\@startsection{section}{1}{0mm}
{\baselineskip\vspace{-.1em}}{\baselineskip\medskip\hrule\medskip}{\raggedright\Large\sf\tmb}}
\renewcommand{\subsection}{\@startsection{subsection}{1}{0mm}
{\baselineskip}{.5\baselineskip\bigskip}{\raggedright\large\hspace{0em}\sf\tmb}}

\makeatother

{}
\definecolor{lightgray}{rgb}{0.99,.99,.99}
\definecolor{gray}{rgb}{0.98,.98,.98}
\definecolor{darkgray}{rgb}{0.8,.8,.8}
\definecolor{darkdarkgray}{rgb}{0.66,.66,.66}
\definecolor{superdarkgray}{rgb}{0.5,0.5,.5}
\definecolor{sand}{rgb}{.79,.83,.85}

\usepackage{dirtytalk}
\usepackage{orcidlink}
\newtheorem{RecommenderRule}{Recommender's Rule}
\newtheorem{AgentsRule}{Agents' Rule}

\usepackage{csquotes}
\usepackage{dirtytalk}
\usepackage{xr}
\usepackage{cleveref}

\title{Can few lines of code change Society? Beyond fack-checking and moderation: how recommender systems toxifies social networking sites}

\begin{document}

\begin{center}
\tmb{\Large {\sc Can few lines of code change Society?}\\ Beyond fack-checking and moderation:\\ how recommender systems toxifies social networking sites}\\
\bigskip

\tmb{\large David Chavalarias$^{1,a,b,*}$, Paul Bouchaud$^{a,b,*}$ et Maziyar Panahi$^{a}$}\\ \smallskip

\bigskip
\end{center}

\begin{quote}
\textbf{Abstract}
As the last few years have seen an increase in online hostility and polarization both, we need to move beyond the fack-checking reflex or the praise for better moderation on social networking sites (SNS) and investigate their impact on social structures and social cohesion. In particular, the role of recommender systems deployed at large scale by digital platforms such as Facebook or Twitter has been overlooked. This paper draws on the literature on cognitive science, digital media, and opinion dynamics to propose a faithful replica of the entanglement between recommender systems, opinion dynamics and users' cognitive biais on SNSs like Twitter that is calibrated over a large scale longitudinal database of tweets from political activists. This model makes it possible to compare the consequences of various recommendation algorithms on the social fabric and to quantify their interaction with some major cognitive bias. In particular, we demonstrate that the recommender systems that seek to solely maximize users' engagement \textit{necessarily} lead to an overexposure of users to negative content (up to 300\% for some of them), a phenomenon called algorithmic negativity bias, to a polarization of the opinion landscape, and to a concentration of social power in the hands of the most toxic users. The latter are more than twice as numerous in the top 1\% of the most influential users than in the overall population. Overall, our findings highlight the urgency to identify harmful implementations of recommender systems to individuals and society in order better regulate their deployment on systemic SNSs.

\end{quote}

\begin{quotation}
\textbf{Significance statement} January 6, 2021 was a shock to democracies. Everything suggests that it was not a fad and social networks played their role. However, to counter the relentless worldwide polarization of public opinion, we need to go beyond  \say{fack-checking} and the moderation of harmful content. This paper studies the role of self-learning recommendation systems on systemic platforms such as Facebook or Twitter and their interaction with users' cognitive bias. We show that their most likely current implementation \textit{necessarily} leads to harmful consequences for individuals and society. Unless BigTech companies prove otherwise, this is not a user behavior problem but a technology problem. This implies that systemic digital platforms currently pose systemic risks to social cohesion. Keys to evidence-based regulation are provided.
\end{quotation}

\bigskip
{\footnotesize 
$^a$ CNRS, Complex Systems Institute of Paris Île-de-France (ISC-PIF), 113 rue Nationale, 75013 Paris, France\\
$^b$ EHESS, Centre d'Analyse et de Mathématique Sociales (CAMS), 75006 Paris, France\\
$^*$ These authors have equally contributed to the study\\
$^1$ Corresponding author : David Chavalarias
}

\section{Rationale}
In January 2018, Facebook announced a change in its \textit{news feed}, a recommender systems which is the main information source of its 2.2 billion users. The aim is to favor content that generates the most engagement: shares, comments, likes, etc. Unfortunately for the public debate, research in psychology shows that this content is, on average, the most negative, a phenomenon called \textit{negativity bias} \citep{rozin_negativity_2001}. The effects of this change are not long in coming. According to leaked internal Facebook documents \citep{hagey_facebook_2021, zubrow_facebook_2021}, exchanges between users have since then become more confrontational and misinformation more widespread. Meanwhile, the political polarization of the users increased due to the platform \citep{allcott_welfare_2020}. These changes were so radical and profound that both journalists and political parties felt forced to \say{skew negative in their communications on Facebook, with the downstream effect of leading them into more extreme policy positions}.

This increase in polarization and hostility in on-line discussions has been observed on other platforms. On Twitter for example, where user's home timeline is by default governed by a recommender system since 2016, the proportion of negative tweets among French political messages raised from 31\% in 2012 to more than 50\% in 2022 \citep{mestre_eric_2022}. It has also been demonstrated \citep{vosoughi_spread_2018} that falsehood diffuses \say{significantly farther, faster, deeper, and more broadly than the trut} on this platform while having the strongest \textit{echo chamber effect} \citep{gaumont_reconstruction_2018}, and consequently the stronger polarization effect.

%access to broadband Internet increases partisan hostility \citep{lelkes_hostile_2017}
Can changing few lines of code of a global recommender system qualitatively change human relationships and society as a whole? To what extent social media recommender systems are changing the structure of online public debates and social group formation processes on a global scale? These are fundamental questions for the sanity of our democracies at a time when polarization in on-line environments is known to spill-over off-line \citep{doherty_partisanship_2016}. Moreover, at a time where countries like the European Union start to regulate the sector of digital services\footnote{Digital Service Act (DSA) https://ec.europa.eu/commission/presscorner/detail/en/ip\_22\_2545 } a scientific answer to these questions is also required to implement evidence-based policies.

Previous studies have explored the societal impact of on-line social networking sites (SNSs) such as the impact of recommender systems on on-line social groups formation \citep{ramaciottimorales,santos_link_2021}, on-line social networks polarization \citep{tokita_polarized_2021} or the impact of networks topologies on opinions dynamics \citep{baumann_modeling_2020}. But the impact of recommender systems on the coupling between opinion dynamics and social network formation is hardly addressed in literature.

%This paper fills this gap analyzing the impact of recommender systems on societies by combining three key aspects: 1) their impact on both opinion dynamics and social networks formation, 2) the feedback loops between personalization and user's cognitive bias and 3) the asymmetry introduced in the inter-personal relations, \textit{i.e.} a recommendation induces a mono-directional social influence, from the recommended account on the target of the recommendation. \com{C'est un point important et je n'ai pas trouvé de papier qui prenne cela en compte.}.

This paper fills this gap and provides a methodological framework that takes into account the entanglement between personalized recommender systems, human cognitive bias, opinion dynamics and social networks evolution. It makes it possible to explore the consequences of various design of recommender systems on the social fabric and to quantify their interaction with some major cognitive bias.

%We propose an integrated model of on-line information systems to analyze the co-evolution of opinion dynamics and social networks formation under different types of recommender systems able to learn from its user's behaviors.

As a case study, we apply this framework to a Twitter-like social network model. We build a state-of-the-art opinion dynamics model and perform an empirical calibration and empirical validation of different components of this framework on a 500M political tweets database, published between 2016 and 2022. Next, we illustrate the impact of recommender systems on society by comparing four differently designed recommender systems based on behavioral, opinion, and network models calibrated on our Twitter data. Perspectives are given to extend this approach to other types of social networks.

This case study highlights the role of human cognitive biases and of the characteristics of new digital environments in the self-reinforcement processes that fragment opinion spaces and distort to a large extent Internet users' perception of reality.

In particular, we demonstrate that, as soon as users have a slight negativity bias,  recommender systems that seek to solely maximize users' engagement lead to an overexposure to negativity, a phenomenon called \textit{algorithmic negativity bias} \citep{chavalarias_toxic_2022} and to a stronger ideological fragmentation of on-line landscapes compared to situations where information circulates in a neutral way.

 %Harmful consequences for the sanity of our democracies resulting from the positive feedback between human flawed cognition and ICT's economic goals.

\section*{Framework Description}
Let's model the generic properties of online social networks in order to study the stylized phenomena associated with some of their key features. The detailed characteristics of SNSs varies from one platform to another but they all have some core features in common:
\begin{enumerate}
\item \textbf{[Publication]} At anytime $t$, a user $i$ can publish a message $m_i^t$,
\item \textbf{[Networking]} Each user $j$ can subscribe to $i$'s information diffusion network $\mathcal{N}_i^d(t)$ (on some social networking sites subscriptions are open, on others they should be agreed by $i$).
\item \textbf{[Information]} Each user $i$ can read the messages produced by the set $\mathcal{N}_i^r(t)$ of accounts they have subscribed to and eventually share them with their own subscribers $\mathcal{N}_i^d(t)$.
\end{enumerate}

Subscription networks between users of a social networking site can be represented by an evolving directed network $\mathcal{N}(t)=\cup_i \{\mathcal{N}_i^d(t) \cup \mathcal{N}_i^r(t)\}=\{s_{ij}\}_{i,j}$ in which an edge $s_{ij}$ exists when the user $j$ has suscribed to $i$'s account (information flows from $i$ to $j$). $\mathcal{N}$ is the backbone of information circulation on such platforms. Its evolution is generally influenced by a \textit{social recommender system} that suggests new \say{friends} to users.

The average number of subscriptions per user being quite high (\textit{e.g.} $>300$ on Facebook, $>700$ on Twitter), most social networking sites implement a \textit{content recommender systems} that helps any user $i$ to find the \say{most relevant} messages among those produced by their social neighborhood $\mathcal{N}_i^r(t)$. On platforms such as Facebook, Twitter, YouTube, LinkedIn, Instagram or TikTok, these content recommender systems take the form of a personalized news feed $\mathcal{F}_i$ that aggregates \say{relevant} messages in a stack. They constitute the main source of information for the users of these platforms (on Youtube the recommender is responsible for 70\% of watch time for exemple\footnote{https://www.cnet.com/tech/services-and-software/youtube-ces-2018-neal-mohan}).
Social recommendation and content systems shape users' opinions through the constraints they place on the global flow of information as well as on the processes of social ties formation. Although most of them are black boxes, we know that these recommender systems learn from the actions of their users according to some very generic objective function.

This being said, in order to model the interaction between human cognition, recommender systems, user's opinions and social network evolution, we have to model three things:
\begin{itemize}
\item \textbf{[Recommender system]} The process of message sorting of news feed by the content recommender system,
\item \textbf{[User's attention, cognitive bias and opinion dynamics]} The user's motivations to read and share a message, their potential cognitive bias, their opinion and its evolution after exposure to a message,
\item \textbf{[Network evolution]} The way users decide to subscribe and unsubscribe to other accounts and the role of the social recommender system in this process.
\end{itemize}

We will include all these elements in a model in discrete time where each time step will correspond to one day of interaction between users. Each of these elements is the object of a research field in its own right, so that it is not a question here of proposing advances on each of these dimensions. We will rather consider the state-of-the-art models for each of them in order to calibrate them on empirical data and study their interactions.
 %Each step, each agent publishes new messages according to its rate of publication $\theta_{i}^p$, theses messages being then displayed to the other users by the recommender.% Agents' properties defined, let's specify model's interaction rules.

\subsection*{The content recommender system}

A content recommender systems has access to a set of users' characteristics, as for example the number of subscribers per user, the list of the accounts they have subscribed to, the number of shares per message, etc., and a set of messages' features, as for example their number of shares or their sentiment, and produces at each time step $t$ and for each user $i$ and ordered list of all messages produced by accounts from $\mathcal{N}^r_i(t)$ to be displayed for reading. %\com{vérifier la temporalité : si je lis un retweet, il a forcément été produit à avant t. Préciser la synchronie / asynchronie entre les actions.}

\subsection*{The Users}
Users are described as entities with an internal state (their "opinion"), some interface with their environments (\textit{e.g.} read some messages) and a repertoire of actions on the environment (publish a message, share a message, subscribe to a user's account, etc.). For simplicity, we assume that the opinion dynamics is solely driven by the interactions among users. We will call this stylized representation of the users "agent".
\subsubsection*{Users' opinions}
We built on the literature of non Bayesian opinion dynamics modeling (see \citep{noorazar_classical_2020} for a review) and assign to each agent $i$ at $t$ an opinion $o_i^t$ in a metric space $\mathcal{O}$, and an opinion update function $\mu_i: \mathcal{O}^2 \rightarrow \mathcal{O}$ that defines its propensity to change its opinion $o_i $ after reading a message that conveys opinion $o_j$ of agent $j$. $\mathcal{O}$ and $\mu$ will be estimated empirically.

\begin{AgentsRule}[Opinions' update]: after sharing agent $j$'s message at time $t$, agent $i$'s opinion is updated according to $o_i^{t+1}\leftarrow \mu_i(o_i^t,o_j^t)$.%, with $\mu_i\in \mathbb{R}$ an update parameter. %Such retweet occurs according to a probability distribution $P(o_j(t)-o_i(t))$, experimentally determined.
		\label{AgentsRule:opinionExp}
\end{AgentsRule}

\subsubsection*{Users' online activity}
At each time step $t$, each agent $i$ publishes $n_i^p(t)$ new messages, assumed to perfectly reflect they view, and shares $n_i^s(t)$ read messages authored by other agents.  $n_i^p(t)$ and $n_i^s(t)$ will be estimated empirically.

\begin{AgentsRule}[reading a message]: at each time step, agent $i$ will \say{scroll} in their feed and randomly stop to read carefully some messages.
	\label{AgentsRule:reading}
\end{AgentsRule}

\noindent Once read, the user may engage with the message:
\begin{AgentsRule}[engagement with a message]: the probability that an agent $i$ shares a message from an agent $j$ (i.e. republish the message identically at the next time step) depends on the difference of opinion $\lvert o_i^t-o_j^t \rvert$.
\label{AgentsRule:engagement}
\end{AgentsRule}

In the literature, different functional forms for the probability of engagement have been proposed such that the exact from should be estimated empirically according to the kind of opinion space that is modeled.

%if the opinion of the author of the message is close enough to the agent's one. In practice, the probability $P(o_i^t,o_j^t)$ that agent $i$ shares a message of $j$ at time $t$ decreases as the difference of opinion $\lvert o_i^t-o_j^t \rvert$ increases.

%The decreasing nature of the probability distribution $P$ implements the \textit{confirmation bias}, the tendency of individuals to prefer messages aligned with their preexisting beliefs over messages that challenge them, as experimentally observed \citep{Knobloch_Westerwick_2017}. %Considering a probability distribution of retweeting a content instead of simple disagreement thresholds ubiquitously used in the literature \citep{Deffuant_2000, Jager_2005}, is motivated by empirical data, discussed in the calibration section.

\subsubsection*{Users' cognitive bias}
%One of the main challenge to anticipate the effect of recommender systems on opinion dynamics is to quantify the feedback loops between human cognitive traits and recommender systems' machinery.
Many cognitive bias are worth to be studied in the perspective of the analysis of recommender systems' impacts. As an illustration, we will focus on two famous bias in psychology: the previously mentioned \textit{confirmation bias} and the \textit{negativity bias} \citep{rozin_negativity_2001, Knobloch_Westerwick_2017, epstein_search_2018} ---the propensity to give more importance to negative piece of information.  Our goal in this example is to evaluate the strength of the \textit{algorithmic negativity bias} \citep{chavalarias_toxic_2022}: the large scale over-exposition to negative contents due to the algorithmic machinery.
%Indeed, one of the key question for research on the impact of social networks on societies is to evaluate to what extent the interaction between recommender systems and user's negativity bias could produce an \textit{algorithmic negativity bias} \citep{chavalarias_toxic_2022} (cf. Fig.~\ref{fig:algo_neg_bias}): a large scale over-exposition to negative contents.

%\begin{figure}[h]
%	\centering
%	\includegraphics[width=0.5\columnwidth]{gallery/algorithmic_negativity_bias.pdf}
%	\caption[Algorithmic negativity bias]{Recommender systems learn from user's engagement that is biased by a \textit{negativity bias}, leading to an  \textit{algorithmic negativity bias}. The \textbf{$+$} signs signal positive feedback and the $ \sim $ a learning effect.}
%	\label{fig:algo_neg_bias}
%\end{figure}

To quantify the \textit{algorithmic negativity bias} effect, we attribute a valence to messages published by the agents, that can be either \say{negative} or \say{positive/neutral}. We thus assign to each agent $i$ a proportion $\nu_i^t$ of negative messages published at $t$ and a propensity $Bn_i$ to interact in a privileged way with negative messages (\textit{negativity bias}). $\nu_i^t$ and $Bn_i$ will be estimated empirically. The negativity bias of our agents is then implemented as a variation of rule \ref{AgentsRule:reading}:

 \begin{AgentsRule}[reading a message with valence]: at each time step, agent $i$ will \say{scroll} in their feed and randomly stop to read carefully some messages. The probability of stopping and read a negative message is $Bn_i$ times higher than for a non-negative message.
 	\label{AgentsRule:readingNeg}
 \end{AgentsRule}

 % \begin{AgentsRule}[reading a negative message]: agent $i$ read negative messages with a probability $p_a\times $.
 % 	\label{AgentsRule:reading_negativity}
 % \end{AgentsRule}

The above defined set of rules allows us to study the feedback loops between the aforementioned cognitive biases and a learning recommender. On the one hand the recommender seeks to maximize the user engagement, on the other hand, the user is more likely to engage with content aligned with their existing belief and/or of negative nature. As a consequence, we can expect the recommender to be more and more biased as it learns users’ bias over time. %Moreover, both the confirmation and negativity biases are reinforced when the user is in a stressful/hostile environment, environment arising from the mere action of the recommender. The need of empirical impresssion information is crucial to explore this last feedback and will be explore in further works.

%Seeking to study the interplay between algorithmic recommendation and human confirmation and negativity biases, each published message is regarded either as negative ---in the psychological sense i.e. unpleasant, offending, harmful--- or not, i.e. neutral or positive. Socio-cognitive studies have indeed revealed that humans are more affected by negative messages than by neutral or positives ones \citep{Knobloch_Westerwick_2017}, hence our choice not to distinguish between neutral and positive messages.

\subsection*{Network evolution}
Opinions co-evolve with interaction networks in a feedback loop. The homophilic nature of human interactions indicates that users tend to interact and form relationships with people who are similar to them \citep{McPherson_2001}, and cut social ties with people who happen to share content that is not aligned with their views. Besides this, SNSs usually suggest new connexions to users via social recommender systems that are most of the time based on structural similarities (\textit{e.g.} mutual friends) \citep{tokita_polarized_2021}.

We will take into account these factors in a parsimonious yet realistic model of link formation and pruning. The network specifications at initialization of our simulations (connectivity, types of agents, etc.) will be determined empirically.

\subsubsection*{Links suppression (Rewiring rule 1)}
Agents score their subscriptions to monitor the interest they have in maintaining them. For every subscription $s_{ji}$ of $i$ to $j$, the disagreement $\delta_{ij}(t) \geq 0 $ of $i$ with the content received through $s_{ji}$ is initialized at $0$ and updated at each time step according to $\delta_{ij}(t+1)=\gamma\times \left(\delta_{ij}(t)+n_{ij}^t \lvert o_i^t - o_j^t\rvert\right)$, with $\gamma<1$ being a daily discount factor and $n_{ij}^t$ the number of messages read by $i$ during time step $t$ that have been authored or relayed by $j$. If $s_{ji}(t)=1$ and the disagreement $\delta_{ij}(t)>\tau$, $i$ will unsubscribe from $j$, \textit{i.e.} $s_{ji}(t+1)=0$. % /!\ souligner par la suite que  \gamma est totalement arbitraire \com{Ce n'est pas simplement $\delta_{ij}(t+1)=\gamma\times \delta_{ij}(t)+n_{ij}^t \lvert o_i^t - o_j^t\rvert$ ?}

%\com{Là ce n'est pas clair : 1) Comme il peut y avoir pour chaque step plusieurs interactions en fonction de la longueur de session et qu'un  message reçu via $s_{ij}$ peut-être un partage d'un message de $k$, l'actualisation devrait s'effectuer même s'il n'y a pas eu de message sur une période de sorte qu'on peut maintenir un lien avec un individu avec qui on est peu aligné mais qui parle peu (Ce serait d'ailleurs une hypothèse à tester) 2) la différence d'opinion à prendre en compte est $\lvert o_i - o_k\rvert.$} :
%
%\begin{Rewire}: If $s_{ji}(t)>1$ and the disagreement $\delta_{ij}(t)>\tau$, $i$ unsubscribes from $j$ ($s_{ji}(t+1)=0$).
%% /!\ souligner par la suite que $\tau$ est totalement arbitraire
%	\label{Rewire:cut_edge}
%\end{Rewire}

$1/\gamma$ is a characteristic time of agents' evolution that is difficult to estimate empirically. It will be set arbitrary to a reasonable value. So will be the $\tau$ which determination would depend on the knowledge of $\gamma$. We have verified that our results do not depend on the precise knowledge of these two parameters.

\subsubsection*{Links formation  (Rewiring rule 2)}
To maintain the connectivity measured empirically, we assume that when an agent breaks an edge with an unaligned user, it starts following a randomly chosen second neighbors (a rewiring mechanism often observed in SNSs \citep{tokita_polarized_2021}).

\section*{Instantiation of a recommender systems : the example of Twitter}
In order to understand the complex relation between the specific choice of a recommender systems and its systemic effects on opinion dynamics and social networks evolution, we apply thereafter the above described framework to the modeling of political opinion dynamics on Twitter. Passing, we find realistic parameter values that could be used to model the impact of other SNSs recommender systems.

At the time of the study, Twitter's data availability, its widespread use ---more than $300$ millions of monthly active users worldwide--- and its predominant role in political communication justify our choice to use it as our experimental field for testing the proposed framework. Moreover, as measured empirically, Twitter is also a digital media where negative contents are more viral than others (see Fig. S18) %Fig.~\ref{fig:neg_by_RT}) 
and where the users themselves are biased towards the production of negative contents (see see Fig. S11) %Fig.~\ref{fig:distribution_negativity}). 
This raises the important question, both for public debate and for the well-being of users, of the extent to which this overflow of negativity is due to Twitter's algorithmic architecture.

Briefly, Twitter is an online social network launched in 2006 allowing its users to exchange publicly 280 characters-long messages that are broadcasted to theirs \say{followers}, users who subscribed the author’s account. Content is displayed to the users on a feed called \textit{Home timeline} according to personalized recommendations. The messages are ranked by a machine learning algorithm predicting the likelihood the user will engage with the tweet. In the following, we will focus on the two main forms of  engagement on Twitter \citep{twitter}: (1) the careful read of a tweet --which often requires a click to expand the content-- (2) the retweet, \textit{i.e.} the fact of republishing the message identically with the mention of its author, without any comment nor modification.

Despite that social influence extents well beyond retweet, empirical studies observed that retweets are more relevant to characterize people's opinion and monitor its evolution, at least in a political context, than, for example, Twitter mentions  \citep{Garimella_2018, Conover_2011}. It is indeed possible to predict with high accuracy the political orientations of political activists from their retweet data only \citep{gaumont_reconstruction_2018}. In what follows, the empirical applications of our framework will focus on retweets networks.

%\N{Note that in addition to show what one's followees has tweeted or retweeted, the recommender will \say{[inject] content from outside the user's immediate neighborhood. A common type of such injected content are tweets liked by someone the individual follows} or \say{promoted tweets, a form of advertisement} \citep{twitter, twitter_political_amplification}.} \com{On peut garder cette phrase si on l'implémente dans le modèle. A voir si c'est le cas.}

\subsection{Choice of a recommender system}
Several leaks as well as official announcements suggest that many social networking sites use the users' engagement maximization as the objective function for their recommender systems. We will thereafter analyze the consequences of such objective functions on the social fabric.

\begin{RecommenderRule}: at each time step, the recommender will rank and display for each agent $i$ a subset of messages from $\mathcal{N}_i^r(t-1)$ according to their probability of being shared, as predicted by the recommender.
	\label{RecommenderRule:display_engagement}
\end{RecommenderRule}
% Effectivement t-1 \com{vérifier (t-1)}

Due its flexibility and efficiency, we implemented this optimization through XGBoost algorithm \citep{XGBoost}.

Recommender systems fulfill their objectives by relying on certain inputs. Modeling such algorithm should thus define some type of data it has access to. The variety of input data used by commercial recommender systems is part of the domain of business secrecy such that little is known about which input data are really used. We will here select two broad categories of data that are likely used by commercial recommenders (cf. \citep{Zhiheng_Xu_2012}, \citep{twitter_political_amplification}) :
\begin{itemize}
	\item \textit{Sentiment analysis}: the negative or neutral nature of a tweets, as well as the proportion of negative content retweeted by the user in the past.
	\item \textit{Popularity assessment}: the popularity of the tweet's author i.e. average number of retweets to its messages, the number of time the message has been retweeted and the frequency at which the user retweets the author.
\end{itemize}

In order to investigate the consequences of the different input features, we will compare three different implementations of the recommender :
\begin{itemize}
	\item \textit{Neg}: use only input data from sentiment analysis,
	\item \textit{Pop}: use only input data from popularity scores,
	\item \textit{PopNeg}: use the combined features of the \textit{Neg} and \textit{Pop} algorithms.
\end{itemize}

To assess the effect of these three implementations of recommender systems on the social fabric, we will compare them to a neutral recommender systems, the reverse-chronological presentation of content, thereafter call \textit{Chrono}. \textit{Chrono} is often referred as non-algorithmic recommendation due to its simplicity. It was briefly implemented by Twitter until the takeover by Elon Musk, from which it is no longer possible to disable the recommendation algorithm.

\section*{Empirical Calibration}

In this section, we fully calibrate our model using empirical data regarding French politics, collected on Twitter in autumn 2021 within the \textit{Politoscope} project \citep{gaumont_reconstruction_2018}, a social macroscope for collective dynamics on Twitter. The \textit{Politoscope} continuously collects since 2016 political tweets about French politics and makes it possible to select subsets of the most active users over any given period.

\subsection*{Network of users' interactions}

While accessing Twitter's graph of followees-followers is possible through Twitter API, such a graph would be misleading if used in our model. Indeed, the content recommender effectively used on Twitter is already well trained, content from someone followed may never be shown to the user, distorting our simulation. To circumvent this limitation, we instead consider the empirical network $\tilde{\mathcal{N}}$ of retweets and quotes combined. Such a network seems indeed to be a reasonable proxy to what is actually shown to the user by the platform. Considering quotes, and not only retweets, allows to include ideologically unaligned content as discussed below. Each of our simulations was initialized over the empirical network $\tilde{\mathcal{N}}$ of interactions over the selected period.

%\N{We employed ForceAtlas2 algorithm to build a metric space from our graph of interaction, while the algorithm is not deterministic, \citep{gaumont_reconstruction_2018} shown that the global topological structure was preserved across multiple spatialization and displayed politically relevant structures, justifying its use.} Having such a metric space allow us to compute the impact of user's exposition to content on their opinion as well as the global impact of different recommender systems on the distribution of the users' opinions.

\subsection*{Calibration of opinion space}
We will henceforth understand the term \say{opinion} as an ideological positioning within the political arena, excluding de-facto political agnostics. Not all candidates having the same digital communication strategy, we will include in what follows only leaders having a significant presence on Twitter during the considered period.

The reconstruction of opinion spaces from SNSs data has been a very active field of research these last several years, with reconstructions in one \citep{Barber__2015,briatte_recovering_2015}, two dimensional spaces \citep{gaumont_reconstruction_2018,chomel_polarization_2022} or even in spaces with variable dimensions \citep{reyero_evolution_2021}. As for retweet networks, retweeting someone on a recurring basis has been demonstrated to be an indicator to some ideological alignment \citep{Garimella_2018, Conover_2011, gaumont_reconstruction_2018}.

\begin{figure}[h!]
	\centering
	\includegraphics[width=\columnwidth]{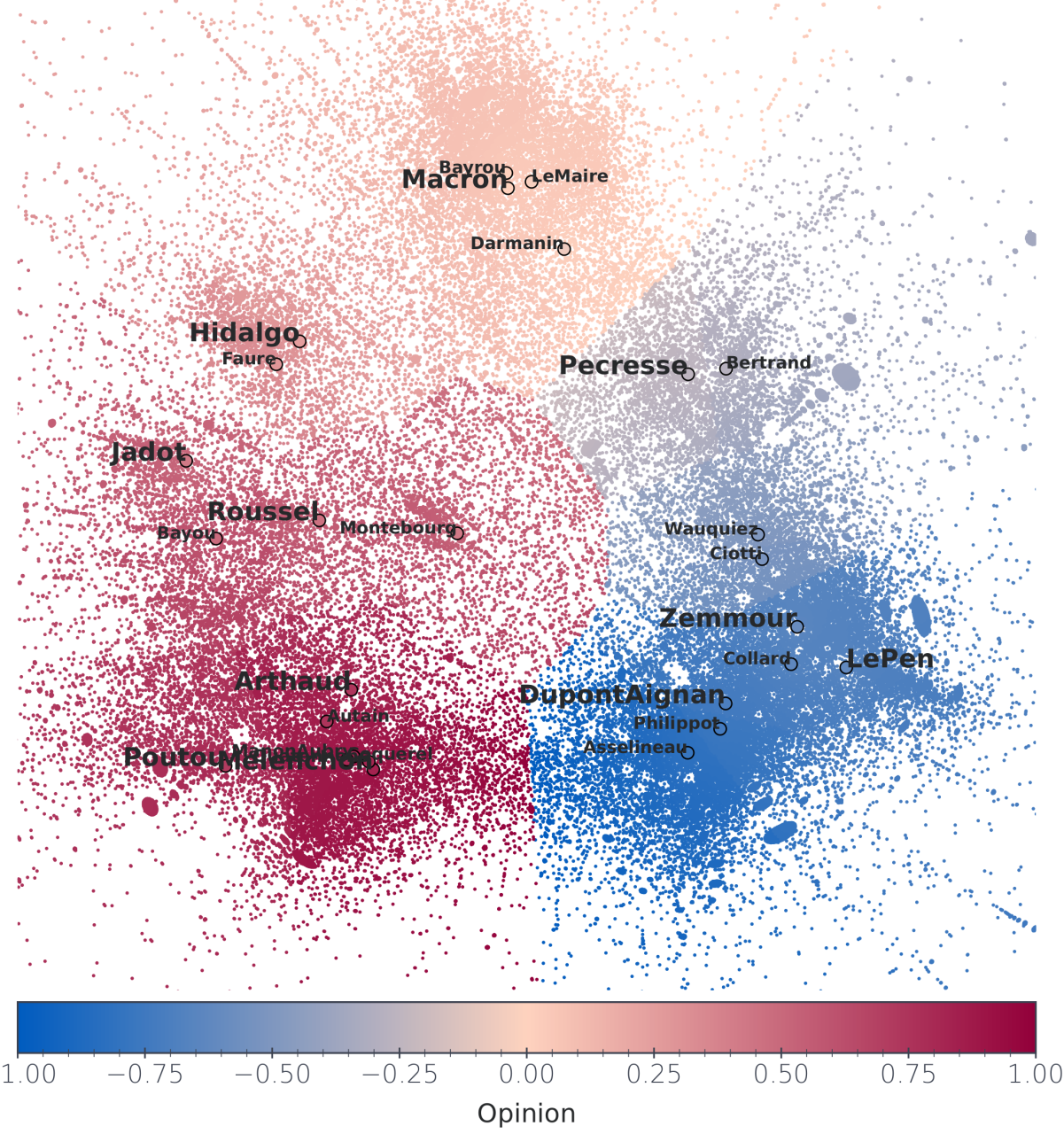}
	\caption[September 2021 Graph RT]{Multi-polar graph of the French pre-electoral political Twittersphere calculated during September 2021. Each node corresponds to a user, colored according to the opinion assigned by the described method. Political leaders are highlighted, in particular the candidates for the 2022 French presidential election.}
	\label{fig:graph_RT_09_2021}
\end{figure}

With a clustering analysis of political retweet graphs, Gaumont et al.  \citep{gaumont_reconstruction_2018} achieved $95\%$ accuracy over opinion's classification, validating the use of the retweet graph for such a study\footnote{We made the same verification on our own dataset and found similar performances.}. The spatialization of the Politoscope retweet graph of autumn 2021 depicts a multi-polar circular political arena (cf. Fig.~\ref{fig:graph_RT_09_2021}) where the relative positions of the political leaders are in adequacy with the publicly depicted political scene. As discussed in SI, we used this spatialization to model the opinion space $\mathcal{O}$ as a circular one dimensional metric space with $o_i \in ]-1,+1]$, making it possible to initialize the opinion of our agents in $\tilde{\mathcal{N}}$ with their empirical estimation, compute the impact of the recommender's suggestion on user's opinion, and determine the global impact of different recommender systems on the distribution of the users' opinions in $\mathcal{O}$.

\subsection*{Calibration of agents' opinion update}

Having a metric space for the opinion space, we can build on the sizable literature on opinion dynamics \citep{Deffuant_2000,Jager_2005,noorazar_classical_2020}. %considering that the gap between two agents' opinions is reduced after their interaction.
Thanks to the full history of user's interaction from our Twitter dataset, and assuming for the sake of simplicity that the functional form of $\mu$, the opinion update function, is the same for all agents, we determined the most likely $\mu$ using symbolic regression. The regression was performed using genetic algorithms \citep{DEAP_JMLR2012} over the set of arithmic and trigonometric functions as well as an implemention of the difference in the periodical opinion space.

This empirical calibration allowed us to identify a linear function of opinion updating $o_i^{t+1}\leftarrow o_i^t + \lambda_i(o_j^t - o_j^t)$  with $\lambda_i \in \mathbb{R}$. Note that this function was already a widely used in the opinion dynamics literature \citep{Deffuant_2000, Jager_2005}.

Because of our lack of information on tweets' impression and given our opinion attribution method, we have decided to simplify the model by assuming that agents change their their opinion, \textit{i.e.} ideological positioning, only when they retweet a message.

Then, we fitted for each agent the opinion update parameter $\lambda_i$, which the absolute value reflects the influenceability of the agent, \textit{i.e.} to what extent will they change opinion when retweeting someone else, using the list of daily messages effectively retweeted by the user (see SI \ref{si:update}). 

Such a fitting leads to a relatively high accuracy, with more than $75\%$ of our final fitted opinions off by less than $0.05$ after 30 iterations (corresponding to end of October, cf. Fig. \ref{fig:absolute_accuracy}). This is less than intra-communities opinion diversity. We should emphasize that the goal of the present work is not to accurately predict the opinion of online social media users, but only to provide a faithful simulation of online users' behavior to study the consequences of algorithmic recommendation. In particular, users' opinion are used within the simulation to determine the probability of retweeting a content, thus being off by $0.05$ in opinion does not alter the behavior of the simulation. The only significant changes of opinion are the larger ones ($\Delta_{op}>0.05$), for which the fitted updates rules leads to a relative error less than $25\%$ for more than $60\%$ of the prediction, and even more accurate for particularly large displacements $\Delta_{op}\in[0.5,1]$ (cf. Fig. \ref{fig:accuracy_relative_30d_forecast}). To confirm the sanity of the used method, we considered other time periods, other graph spatialization settings and forecast the opinions one month (November) after the fitting, obtaining similar accuracy, as discussed in supplementary text.

\subsection*{Calibration of agents' activities}

In absence of information specifying which messages are displayed on users' screens, we hypothesize that users read messages until they reach their daily number of retweets or when they read all the messages from $\mathcal{N}_i^r(t-1)$.
We identified the $110k$ most active users over the period of autumn 2021, get their political tweets and estimated their publication behaviors. %\com{@Paul : il faut mettre le graph correspondant anonymisé dans les données associées au papier DE et ajouter ici le nombre exact de tweets traités}.
The number of daily posted tweets ($\tilde{n_i^p(t)}$, original publications) and retweets ($\tilde{n_i^s(t)}$, shared publications) were exponentially distributed at the individual level (as already observed in \citep{perra_activity_2012,baumann_modeling_2020}). At the population level, the empirical exponential scales $\tilde{\theta_{i}^p}$ and $\tilde{\theta_{i}^s}$ for the different users were distributed according the distribution displayed on Fig. S1.
%\ref{fig:agents_activities}. 
We build on these empirical observations to set the number of tweets and retweets of agent $i$ in $\tilde{\mathcal{N}}$  as independently drawn from two exponential distributions of empirically determined rates $\tilde{\theta_{i}^p}$ and $\tilde{\theta_{i}^s}$ respectively.

 %We in addition notice a strong diversity of behaviors, some user publishing almost only retweets, other only new tweets, as depicted by the distribution of users ratio $\theta_{RT}^i/\theta_{TW}^i$.

% First, at the individual level the distributions of number of daily posted tweets and retweets are exponentially distributed. At the population level, the exponential scales $\theta_{RT}^i$, $\theta_{TW}^i$ for the different users are distributed according to a power-law as display on \ref{fig:activity}. We in addition notice a strong diversity of behaviors, some user publishing almost only retweets, other only new tweets, as depicted by the distribution of users ratio $\theta_{RT}^i/\theta_{TW}^i$.

\subsection*{Latitude of acceptance}

Once the opinions assigned, we determined the distribution of difference of opinion $\Delta_{op}$ between a user and the authors of retweeted messages. In order to cancel the bias in the representation made by the platform \citep{twitter_political_amplification}, as well as taking into account the different sizes and positions of the communities, we had to renormalize the distribution of difference of opinion as observed from the retweets with the patterns of publication on quoted tweets (cf. S1.4).%\ref{SI:acceptance}).

\begin{figure}[h!]
	\centering
	\includegraphics[width=\columnwidth]{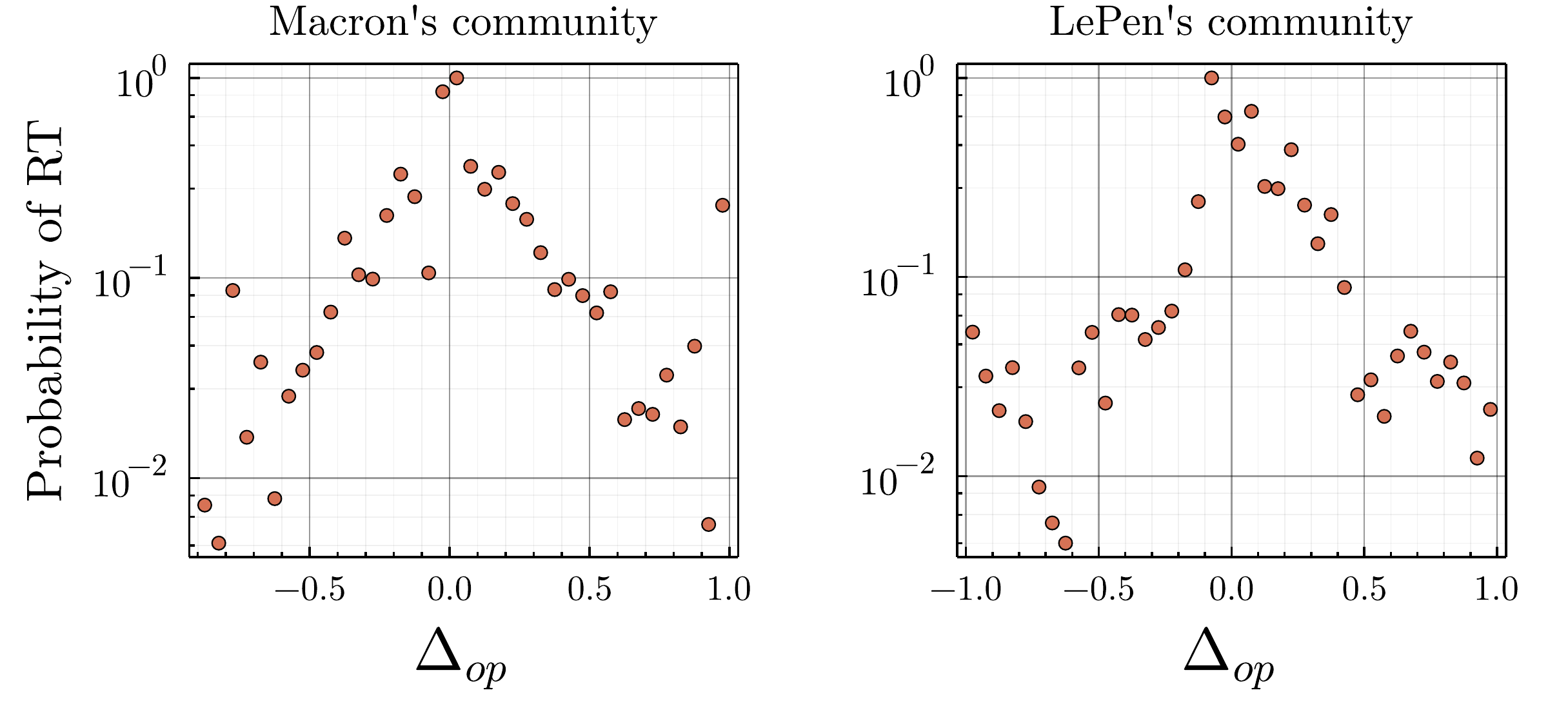}
	\caption[diff-op-RT-quote]{Estimated probability for a user, in the ideological neighborhood of Emmanuel Macron or Marine Le Pen, to retweet a read message according to the different of opinion with its author, $\Delta_{op}=o_{reader}-o_{sender}$, considering periodic boundary conditions. We renormalize such that a perfectly aligned message is retweeted with certainty.}
	\label{fig:diff_op_RT_quote}
\end{figure}

One notices on Fig.~\ref{fig:diff_op_RT_quote}, that the probability of retweeting a message decays roughly exponentially as the difference of opinion increases, with some refinement revealing political strategies. The asymmetry of the distributions illustrates how Emmanuel Macron community (center) tends to retweet content even further right than further left, while the opposite effect is noticed for left-wing leader Jean-Luc Melenchon community (cf. SI S3) %Fig.~\ref{fig:acceptanceTrio}). 
After having determined such distributions for the whole range of opinions, we assigned to our simulated agent the distributions associated to their initial opinions.

\subsection*{Negativity}

To calibrate negativity-related properties of our model, we performed a sentiment analysis on $190k$ French political tweets exchanged by $500$ unique users during October 2021. This analysis has been performed using the French version of the Bi-directional Encoders for Transformers, CamemBERT \citep{Martin_2020}, fine-tuned on French Tweets. We then assigned to our agents an intrinsic negativity $\nu_i$, the proportion of negative content published, drawn from the empirical distribution in function of their initial opinion, as well as a negativity bias, as discussed in supplementary materials.

%\begin{figure}
  %  \centering
 %   \includegraphics[width=\columnwidth]{images/calibration/accuracy_relative.pdf}
  %  \caption{Cumulative accuracy of the fitting procedure in function of the relative error for various opinion displacements}
 %   \label{fig:fit_mu_rel}
%\end{figure}
\subsection*{Evaluation of recommenders' effects}

In order to characterize the behavior of our agent-based model we hereby introduced metrics of particular interest:
\begin{description}
	\item[Algorithmic negativity bias $\Gamma$:] this is the negativity over-exposure generated by the recommender system defined as the ratio between the negativity in the perceived environment ---the content of the timeline--- and the negativity in the \say{real environment} \textit{i.e.} in one's in-neighborhood $\mathcal{N}_i^r$.
	%\item[Popularity:] $p_i$ of an agent $i$, corresponds to the average number of retweets by published message.
\end{description}
To further explore the model, we perform a community detection on the resulting retweets graph using Leiden algorithm \citep{Traag_2019}, an improvement guaranteeing connected communities over the usual Louvain method. Once performed we examine:
\begin{description}
	%\item[Number $K$ of clusters/communities:] in the digraph of follow/RT
	\item[Newman's modularity $\mathcal{Q}$ ]\citep{Leicht_2008}: assessing the density of connections within a community.

	\item[Diversity within a community $\sigma_{X}^{intra}$:] % $$\sigma_{X}^{intra}=\frac{\sum_{k=1}^K \sqrt{\frac{1}{\lvert \mathcal{N}_k\rvert}\sum_{j \in \mathcal{N}_k }\left(X_j-\overline{X_k}\right)^2}}{K \sqrt{\frac{1}{\lvert \mathcal{N} \rvert}\sum_{j \in \mathcal{N} }\left(X_j-\overline{X}\right)^2}}$$ with $\mathcal{N}_k$ a cluster subgraph and $\overline{X_k}$ the average of observable
	the standard deviation of an observable, such as the opinion, the intrinsic or perceived negativity, within a given cluster, normalized by the standard deviation of the observable within the overall population, averaged over the clusters.%. Assessing how, within a community, agents are aligned in opinion/negativity with respect to the overall population.
	\item[Diversity between communities:
	$\sigma_{X}^{inter}$:] the standard deviation of clusters' observable --such as average opinion, intrinsic or perceived negativity-- normalized by the diversity among the whole population, assessing the diversity of the different communities with respect to the overall population.
	%$$\sigma_{X}^{inter}=\frac{1}{ \sqrt{\frac{1}{N}\sum_{j \in G }(X_j-\overline{X})^2}} \sqrt{\frac{1}{K}\sum_{k=1}^K(X_k-\overline{X})^2}$$
\end{description}

\section*{Results}

\subsection*{Assessing the impact of recommenders on negativity and opinion polarization with empirical networks}
The model was initialized on real data with \textit{all} parameters but two, which we know have no impact on the results, being empirically calibrated. We have simulated one month of interactions to estimate the activity and opinion evolution of each account in the real dataset, and analyzed the aforepresented metrics. As for the account's activity prediction, our framework being stochastic, none of the four recommender systems where able to predict low intensity interactions $\leq10$ retweets/month), overestimating small weights with respect to the real distribution (see Fig.~S1).%\ref{fig:agents_activities}).
Nevertheless, for larger weights ($> 10$ retweets/month), \textit{PopNeg} faithfully matches the empirical distribution, while \textit{Pop} overestimates large weights, as one may expect, and \textit{Chrono} underestimates them.

\begin{figure}[h!]
    \centering
    \includegraphics[width=\columnwidth]{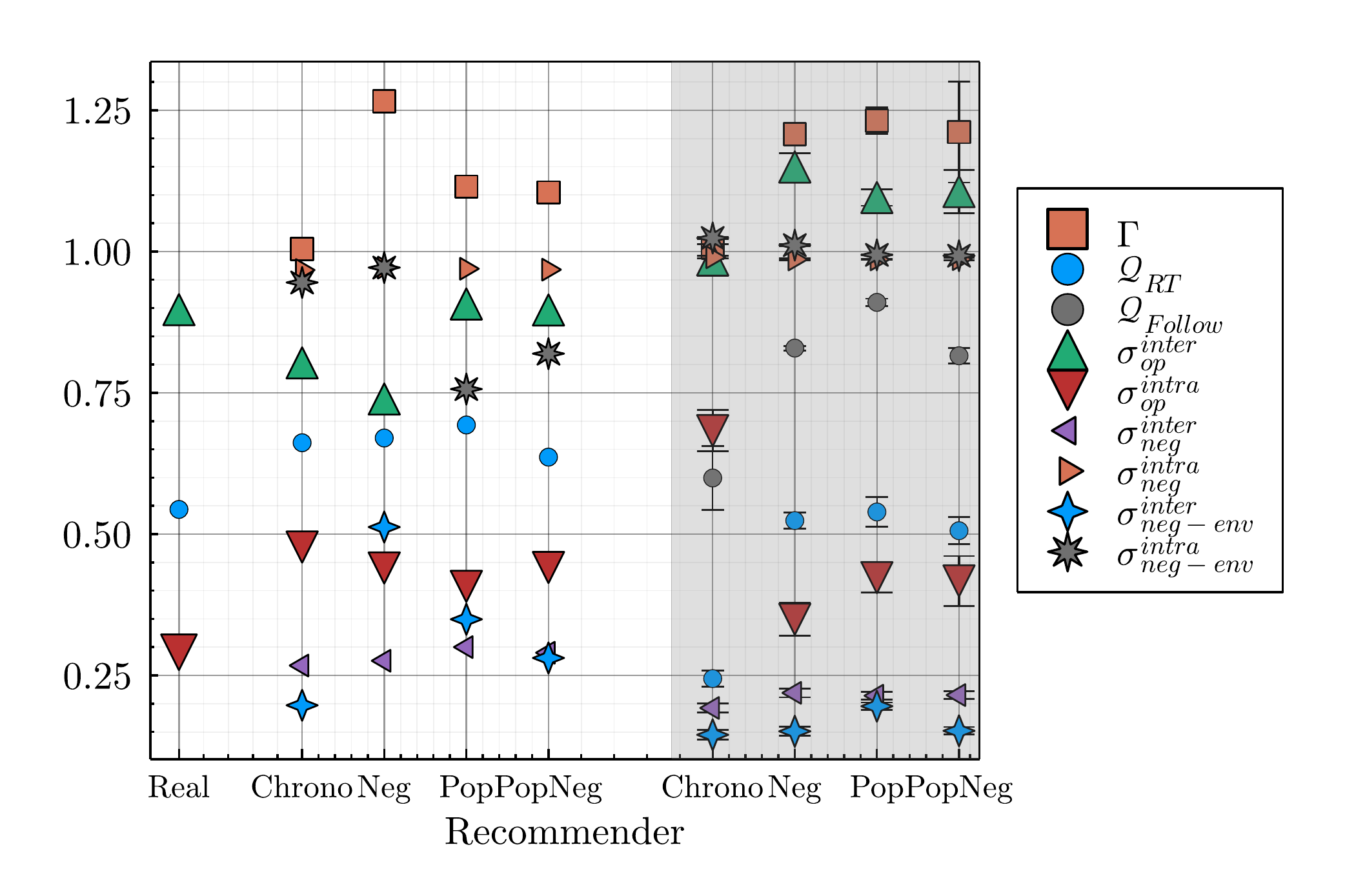}
    \caption{Metrics comparison between the four recommenders and the empirically observed data. Simulation initialized on real graph (white area) and on synthetic graph (grey area), the error bars in the latter case correspond to the standard deviations over 10 repetitions, starting from the same synthetic inputs}
    \label{fig:metrics_graph}
\end{figure}

As displayed on figure \ref{fig:metrics_graph}, the overexposure to negativity $\Gamma$ is non-existent in chronological mode, as expected, while the three algorithmic recommenders lead users to be overexposed to negative content. The \textit{Neg} recommender, solely based on negativity, leads to the highest overexposure, users are shown on average $26\%$ more negative content that what they would have in the neutral \textit{Chrono} mode.

\begin{figure}[h!]
    \centering
    \includegraphics[width=0.7\columnwidth]{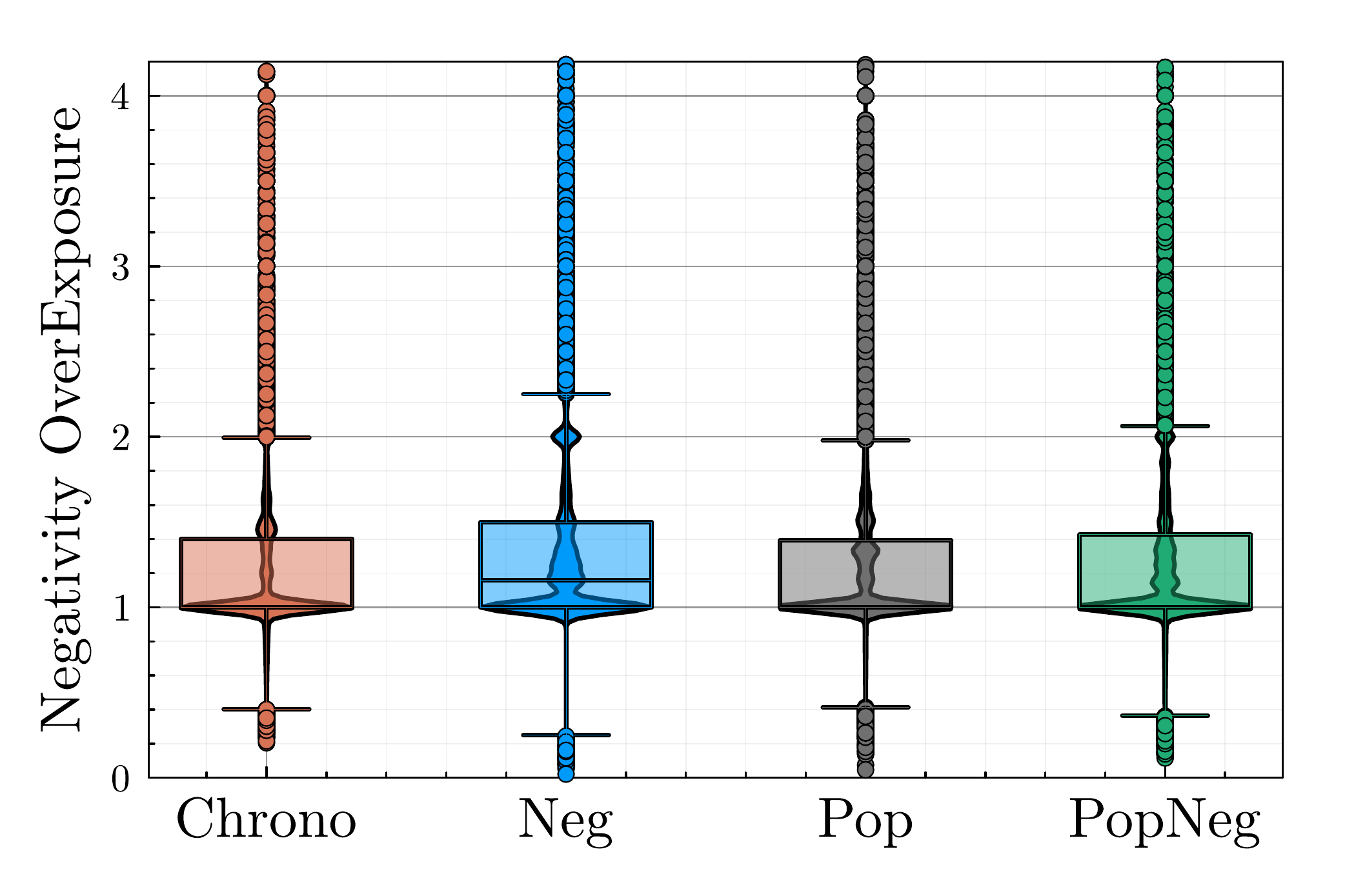}
    \caption{Distribution of the overexposure to negativity within the population for the four recommenders. For clarity, the distribution is truncated at a overexposure of 3.5, the truncated tails represents 3.1\%, 2.6\%, 2.6\% and 3.6\% of the total distribution, for \textit{Chrono,Neg,Pop,PopNeg} respectively}
    \label{fig:violin_overneg}
\end{figure}

Within the population, the overexposure to negativity is extremely diverse, as depicted in Fig.~\ref{fig:violin_overneg}, with some users  experiencing an algorithmic negativity bias corresponding to an overexposure of more than $300\%$. This happens even to users with a large neighborhood and to users without any negative bias. For user with a small number of friends (less than 10), we notice a small ($r=0.02$) but significant ($p<10^{-7}$) correlation between the number of in-neighbors and the negativity overexposure. Indeed, as the number of friends increases, so does the size of the pool of message from which the recommender is picking from, allowing it to select the most engaging messages (that are most of the time the most negatives ones), leading to a higher negativity overexposure. Such results are a direct consequence of the feedback loop between human negativity bias and the engagement maximization goal hard-coded within the recommender.

The large variations in the level of negativity overexposure at the individual level are important to note since from an individual perspective, it can plunge users into toxic environments, disconnected from reality, which can potentially have detrimental consequences on their mental health and social relationships, as documented, for example, during the COVID-19 pandemic \citep{levinsson_conspiracy_2021}.

The diversity of opinion depends of the recommender system, pointing to another harmful consequences for online sanity. Indeed, while \textit{Chrono} and \textit{Neg} lead to the same $\sigma^{inter/intra}_{op}$, the two social modes, namely \textit{Pop} \& \textit{PopNeg}, result into a higher fragmentation of the social fabric. The average diversity of opinion within clusters, $\sigma^{intra}_{op}$, is poorer ---but not as poor as empirically observed---, and the different clusters are centered around different opinions ---higher $\sigma^{inter}_{op}$, close to the empirically observed one.

In contrast, the modularity $\mathcal{Q}_{RT}$ of the retweet graph revealed to be independent of the recommender, as well as the diversity of negativity within and between clusters $\sigma^{inter/intra}_{neg}$, the graph structure being strongly constrained from the initialization

%\com{J'ai l'impression qu'on peut même dire cela en toute généralité puisque cela semble aussi indépendant sur les graphes synthétiques bien que certains valeurs ne soient pas claires en raison de la supperposition des markets.}.
% La modularite du graph de RT est, pour les graphes synthetic, dependent du recommender, at least between chrono mode and "true" algo recommandation.

By looking at recommender features importance, we notice that the frequency of past interactions between the user and the author, is by far the most informative feature, another illustration of human confirmation bias, reinforced by the recommender. Similarly, the different clusters are, in these social modes, less diverse in perceived negativity $\sigma^{intra}_{neg-env}$. The unequal perceived negativity, may partially justify the difference of acceptance latitude for the different opinion, but further experiment considering impressions information would be needed to assess the relation between perceived negativity and confirmation bias.

\subsection*{Assessing the impact of recommenders on the formation of social groups with synthetic networks}
The previously considered empirical data are the product of years of evolution, shaped by the platform's recommenders. Thus by initializing the network of interactions with a real network, we miss most of the impact of the different recommender systems’ on social networks formation. In order to further investigate the consequences of the recommender on the social fabric, we hereby consider randomly initialized networks and analyze their evolution\footnote{Source code available on \citep{DVN/WRFMMU_2023}}. We drawn for $25k$ agents the properties from the empirical distributions and considered an initial network of follow generated through the Barabási Albert model. Such networks do not aim to realistically mimics all real social networks features but only to provide a zero-th order starting point to illustrate the different consequences of the recommender.

The probability of retweeting a read message is set to decays exponentially with the difference of opinion with a mean of $0.2$, to roughly match the empirical one, without specifying it too strongly to French political strategies. The empirical determination of $\tau$ being impossible, without having access to what messages is shown to the users on a long time period, we arbitrarily fixed it $0.5$ with a time discount factor of $0.9$, corresponding to a time-scale of 10 days --- by considering alternatives values, the qualitative results discussed below remains.

A sensitivity analysis of the agents' negativity bias in synthetic networks also demonstrates that the algorithmic negativity bias phenomena ($\Gamma>1$) appears as soon as agents have some negativity bias ; and its intensity is almost independent of the strength of agents' own negativity biases (see  Fig.~S16).
%\ref{fig:overexposure_vs_negbias}). 
As long as the users favor negative content over neutral ones, recommender systems based on engagement will lead with certainty to an overexposure to negativity.

\begin{figure}[h]
	\centering
	\includegraphics[width=1\columnwidth]{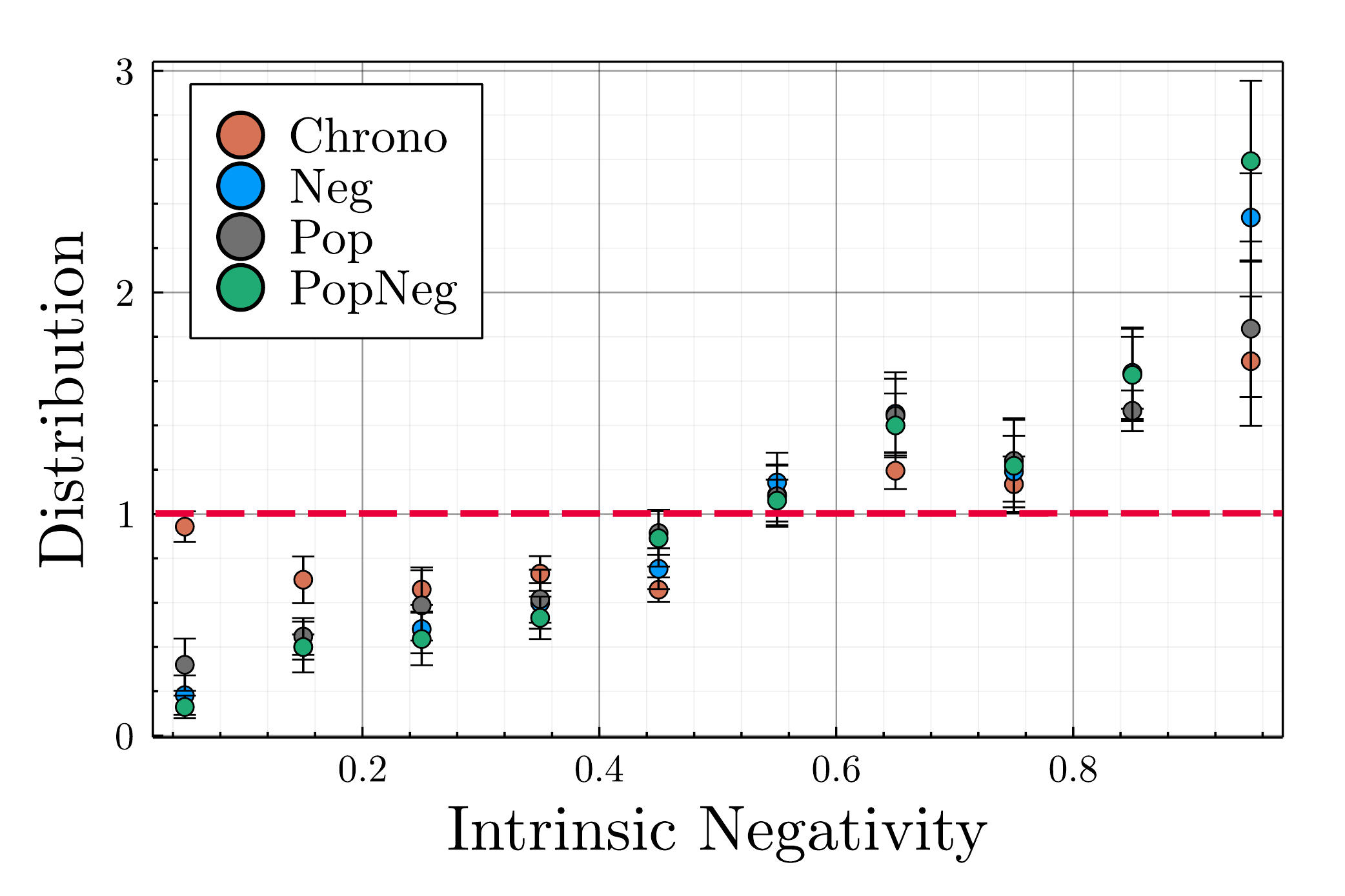}
	\caption[Over-representation of negative agent among popular ones]{Over-representation of negative agent among the 1\% most popular agents compared to the overall population. Analysis performed after two months of simulation, the error bars correspond to the standard deviations over 10 repetitions, starting from the same synthetic inputs.}
	\label{fig:over_representation_neg_popular}
\end{figure}

Starting with an unconstrained random network $\mathcal{N}$ allows the full expression of recommender actions and makes it possible to check that the proposed model for network evolution is compatible with what is observed empirically (see Fig.~S12 % \ref{fig:real_vs_synthetic}
 for an example). As depicted on Fig.~\ref{fig:metrics_graph}, after two months of simulated evolution, the modularity of the retweet and follow networks significantly increases with algorithmic recommendation in respect with a neutral presentation of content, in \textit{Chrono}, as well as the ideological fragmentation or the overexposure to negativity.

The algorithmic negativity bias does not only impacts the information environment of the agents toward more toxic environements, is also impacts the structure of \textit{social power} in the population, defined as the capacity of an agent to influence the public debate \citep{jia_opinion_2015}.
Fig.~S13 
%\ref{fig:negativity_unbalance} 
 displays the intrinsic negativity unbalance between the overall population and the top $1\%$ agents receiving the most retweet by tweet while Fig.~S14 
 %\ref{fig:over_representation_neg_popular} 
 shows the proportion of negative agents in function of the most popular quantile for \textit{Neg} algorithm. This analysis clearly demonstrates that the amplification of individual negativity bias by engagement-optimizing recommendation algorithms leads to a concentration of online social power in the hands of the most negative users.

For example, while agents publishing negative content half of the time are faithfully represented among the most popular, the users publishing no negative content are absent from the most popular ones for the three algorithmic recommenders. Frighteningly, agents publishing systematically negative content are more than twice as numerous among the most popular than in the overall population; the two recommenders considering the negativity of the message, namely \textit{Neg} and \textit{PopNeg}, leading to the highest over-representation.

It is also noteworthy that despite being neutral in its selection, \textit{Chrono} nevertheless leads to a significant unfair representation as a consequence of individual  negativity bias. Even in a neutral mode, the users will more likely read negative content and hence retweet it, increasing its author popularity.

\section*{Discussion}

On January 6, 2021, a crowd convinced that the election was stolen stormed the Capitol in Washington, D.C. Whatever the extent to which this event can be attributed to misinformation about the electoral process, it is clear that it was not a fad: one year after Jan. 6 \say{52\% of Trump voters, as well as 41\% of Joe Biden voters, somewhat agree or strongly agree that it is time to cut the country in half} \citep{politics_new_2021} while a late 2020 survey concluded that \say{Americans have rarely been as polarized as they are today} \citep{dimock_america_2020}. In order to remedy this situation of extreme polarization of public opinion, which tends to be reproduced in other countries such as Brazil, the United Kingdom or Italy, we must go beyond the reflex of \say{fack-checking} and the praise of better moderation of harmful content in online social networks.

As pointed out by other studies using complementary approaches to ours \citep{ceylan_sharing_2023, tornberg_how_2022}, we must acknowledge the impact of SNSs on social structures and in particular in the amplification of polarization and hostility among groups. It is not only a phenomenon that affects the general public, the entire information ecosystem is at risk. After Facebook changed its algorithm in 2018 to favor \say{meaningful social interactions}, \say{company researchers discovered that publishers and political parties were reorienting their posts toward outrage and sensationalism} and internal memo mentionned that \say{misinformation, toxicity, and violent content are inordinately prevalent among reshares} \citep{hagey_facebook_2021}. 

At a time when states are thinking about regulating large social networking sites (SNSs), it is more necessary than ever to have models to quantify their effects on society. In this article, thanks to the modeling of social networks as complex systems and the calibration of the models using big data, we could give hints about what is really going on under the hood.

Using a large scale longitudinal database of tweets from political activists \citep{gaumont_reconstruction_2018}, we have built and calibrated an agent-based model able to simulate the behavior of users on Twitter, some of their cognitive biases and the evolution of their political opinion under the influence of recommender systems. Among other things, we have empirically estimated  parameters common to many models of opinion dynamics that were previously arbitrarily defined --like the widespreadly used opinion update rule Agents Rule \ref{AgentsRule:opinionExp}. We also went beyond commonly adopted assumptions, such as a fixed threshold of ideological disagreement for engaging in an interaction, by considering interaction propabilities and estimating their law.

Thanks to this calibrated model, we could compare the consequences of various recommendation algorithms on the social fabric and to quantify  their interaction with some major cognitive bias. In particular, we demonstrated that the recommender systems that seek to solely maximize users' engagement \textit{necessarily} lead to an overexposure of users to negative content, a phenomenon called \textit{algorithmic negativity bias} \citep{chavalarias_toxic_2022} and to an ideological fragmentation and polarization of the online opinion landscape. 

The important point is that these consequences of the way recommender systems are currently implemented, which are harmful to individuals and society, are not necessarily intentional, they only results from the positive feedback between human flawed cognition and SNSs' economic goals. As most of these platforms have become systemic due to their size, their unregulated pursuit of profitability poses systemic societal risks both to their users and to the sanity of our democracies.

Policy makers are increasingly aware of the threats posed to our democracies by the current implementation of SNSs but lack the keys to regulate this sector. Modeling SNSs and their effect on individuals an social groups with an interdisciplinary approach can give some of those keys. 

For example, we have shown that when a self-learning algorithm is used to recommend content, feeding it with measures of user or content popularity leads to an increase in the overall toxicity of the platform for individuals and the collectives they form. It also lead to a concentration of the social power in the hand of the most toxic accounts. This means that regulators should focus on the types of data that feed into recommender systems and potentially outlaw some. These kind of findings could help to identify when and where business secrecy, which is so often brandished when platforms are asked to cooperate with public bodies, must be relaxed in the context of their regulation.

On the other hand, platforms can take steps to mitigate negativity bias at the user level and prevent it from becoming algorithmic negativity bias and spreading to the collective level.

Studies of the effect of recommender algorithms are an emerging field in academia that should be supported by the relevant authorities in order to identify, in all independence, the right regulatory levers and implement an evidence-based policy. Needless to say, this will require greater openness of SNSs data towards academia\footnote{Let us remind that Twitter has been one of the sole large SNSs to make some of its data available and was considering ending this data openness at the time this article was written.}. Some of the empirical calibration made on Twitter in this study, like the opinion update rule or the reshare behavior, could be usedful to model other platforms like Facebook, but nothing compares to an empirical calibration on the native data of a platform 

In conclusion, it is not enough to point to malicious users who produce toxic content and call for better moderation. We need to further study the effects, at the individual and collective levels, of large-scale deployment of recommender systems by major technology companies and assess their potential harm. Science shall contribute to evidence-based policymaking by modeling the impact of these platforms on the social fabric. Democracy is at stake.

\subsection*{Aknowledgments} 
We are thankful to Victor Chomel for the fruitful discussions, as well as Jeanne Bruneau--Bongard for her careful proofread of the first draft of the present article. This work was supported by a grant from the \say{Fondation CFM pour la Recherche}, as well as the support of the Complex Systems Institute of Paris Île-de-France and the Region Île-de-France.

% Bibliography
\bibliography{main}

\end{document}